\documentclass[fleqn,usenatbib]{mnras}

\usepackage{newtxtext,newtxmath}
\usepackage{amsmath}
\usepackage[T1]{fontenc}

\DeclareRobustCommand{\VAN}[3]{#2}
\let\VANthebibliography\thebibliography
\def\thebibliography{\DeclareRobustCommand{\VAN}[3]{##3}\VANthebibliography}

\usepackage{graphicx}	% Including figure files
\usepackage{amsmath}	% Advanced maths commands
\newcommand\project[1]{\textsl{#1}}
\newcommand\software[1]{\texttt{#1}}
\newcommand\corot{\project{CoRoT}}
\newcommand\kepler{\project{Kepler}}

\newcommand\MESA{\software{MESA}}
\newcommand\ADIPLS{\software{ADIPLS}}
\newcommand\pymc{\software{PyMC3}}

\newcommand\thestar{KIC~12508433}

\newcommand{\Revision}[1]{#1}
\newcommand{\TW}[1]{#1}

\newcommand{\Add}[1]{#1}
\newcommand{\Del}[1]{}

\usepackage{xcolor} 
\definecolor{aureolin}{rgb}{0.99, 0.93, 0.0}

\usepackage[math]{cellspace}
\title[Internal rot. profile constraints from surface rot.]{Constraining the Rotation Profile in a Low-Luminosity Subgiant with a Surface Rotation Measurement}

\author[Tanner A. Wilson et al.]{
Tanner A. Wilson,$^{1,2}$\thanks{Corresponding Author E-mail: tanner.wilson@monash.edu}
Andrew R. Casey,$^{1,2}$
Ilya Mandel,$^{1,3}$
Warrick H. Ball,$^{4}$
Earl P. Bellinger,$^{5,6}$
\newauthor
and Guy Davies $^{4,6}$
\\
$^{1}$School of Physics \& Astronomy, Monash University, Victoria, Australia\\
$^{2}$Center of Excellence for Astrophysics in Three Dimensions (ASTRO-3D)\\
$^{3}$The ARC Center of Excellence for Gravitational Wave Discovery -- OzGrav, Australia\\
$^{4}$School of Physics and Astronomy, University of Birmingham, Edgbaston, UK\\
$^{5}$Max Planck Institute for Astrophysics, Garching, Germany\\
$^{6}$Stellar Astrophysics Centre (SAC), Department of Physics and Astronomy, Aarhus University, Aarhus C, Denmark
}

\date{Accepted XXX. Received YYY; in original form ZZZ}

\pubyear{2023}

\begin{document}
\label{firstpage}
\pagerange{\pageref{firstpage}--\pageref{lastpage}}
\maketitle

% Abstract of the paper
\begin{abstract}
Rotationally-induced mode splitting frequencies of low-luminosity subgiants suggest that angular momentum transport mechanisms are 1-2 orders of magnitude more efficient \Revision{in these stars} than predicted by theory. Constraints on the rotation profile of low-luminosity subgiants could be used to identify the dominant mechanism for angular momentum transport. We develop a forward model for the rotation profile given observed rotational splittings, assuming a step-like rotation profile. We identify a consistent degeneracy between the position of the profile discontinuity and the surface rotation rate. We perform mock experiments that show the discontinuity position can be better constrained with a prior on the surface rotation rate, which is informed by star spot modulations. We finally apply this approach to \thestar, a well-studied low-luminosity subgiant, as an example case. With the observed surface rotation prior, we obtain a factor of two increase in precision of the position of strong rotation gradient. We recover the literature values of the core and surface rotation rates and find the highest support for a discontinuity in the radiative zone. \TW{Auxiliary measurements of }surface rotation could substantially improve inferences on the rotation profile of low-luminosity subgiants with already available data.

%If the surface rotation rate is known to within $\approx10\%$ from star spot modulations, this prior can substantially improve inferences of the rotation profile in low-luminosity subgiants.
\end{abstract}

% Select between one and six entries from the list of approved keywords.
% Don't make up new ones.

%%%%%%%%%%%%%%%%%%%%%%%%%%%%%%%%%%%%%%%%%%%%%%%%%%

%%%%%%%%%%%%%%%%% BODY OF PAPER %%%%%%%%%%%%%%%%%%

\section{Introduction}
\label{sec:intro}

All stars rotate. The rotation rate and distribution of angular momentum throughout a star evolves with time. The effects of rotation on the structure and evolution of a star are substantial \citep[e.g.][]{heger_presupernova_1998,maeder_evolution_2000}, and accurate prescriptions of rotation in stellar models are important to reproduce observations, particularly asteroseismic measurements of oscillation modes.

Low-mass subgiants and low-luminosity red giants pulsate in mixed modes, which are sensitive to structure in both the core and envelope. Mixed modes result from the near-surface convection, which drives the oscillation modes to amplitudes that are detectable in space-based photometry missions like \corot\ \citep{baglin_corot_2003} and \kepler\ \citep{borucki_kepler_2010}. Rotation lifts the degeneracy of oscillation modes of the same angular degree and azimuthal order. The change to the frequency of rotationally split modes is related to the rotation profile. For this reason, measuring rotational splittings allows us to constrain the rotation profile.
%\begin{figure*}
%\centering
%    \includegraphics[width=0.8\textwidth]{hr.pdf}
%    \caption{Hertzsprung-Russell diagram showing effective temperature versus log luminosity of \kepler{} stars from \citep{mathur_revised_2017}. Subgiants and low-luminosity red giants studied in \citep{deheuvels_seismic_2014} are indicated in black, with \thestar{} indicated by a cross. Black curves indicate tracks of non-rotating models (\citep{choi_mesa_2016,dotter_mesa_2016}).}
%    \label{fig:hr}
%\end{figure*}

Current measurements of rotational splittings place low precision constraints on the core and surface rotation rates ($\approx$10\% and 30\% respectively \Revision{\citep[e.g.][]{deheuvels_seismic_2014,fellay_asteroseismology_2021}}) and have little capability to constrain the shape of the rotation profile connecting the core and the surface. These observations, however, present some notable results.
Subgiants demonstrate solid-body rotation early in their transition off the main-sequence (MS), like their MS counterparts \citep{deheuvels_seismic_2020,noll_probing_2021}. However, later in their post-MS evolution, the cores of more evolved subgiants and low-luminosity red giant branch (RGB) stars rotate much faster than their envelope. The core-to-surface rotation ratio can grow to $\approx$20 for stars leaving the subgiant phase \citep{deheuvels_seismic_2014, gehan_core_2018, eggenberger_asteroseismology_2019}.

Stellar models predict differential rotation between the core and the surface to be two to three orders of magnitude greater than observations suggest \citep{eggenberger_angular_2012, cantiello_angular_2014}. The surface rotation rates of white dwarfs agree well with the core rotation rates of RGB stars \citep{gough_glimpses_2015,hermes_deep_2017}, suggesting an angular momentum transport mechanism that is much more efficient throughout the first ascent of the RGB, and nowhere else \citep{eggenberger_angular_2012,marques_seismic_2013,ceillier_understanding_2013, fuller_asteroseismology_2015,spada_angular_2016,ouazzani_gamma_2018}.

% Several mechanisms of additional angular momentum transport could explain the disparity between models and observations. These include strong core-envelope coupling of first ascent RGB stars \citep{tayar_implications_2013}, magneto-rotational instabilities in the form of the Tayler-Spruit (TS) dynamo \citep{cantiello_angular_2014} and the modified TS-dynamo \citep{fuller_slowing_2019}, internal gravity waves \citep{fuller_angular_2014}, as well as angular momentum transport through mixed modes \citep{belkacem_angular_2015}. \citet{spada_angular_2016} parameterised the efficiency of angular momentum transport using a core-surface ratio diffusion coefficient. Their model can reproduce both subgiant and RGB measurements, but it does not identify a physical mechanism. These treatments of angular momentum transport are work-in-progress, and no mechanism can currently describe the rotation rates along the post-MS evolution pathway.

% Inverting a stellar rotation profile given measured rotational splittings is an ill-posed problem. State-of-the-art methods involve the use of either linear inversion techniques or forward modelling. Examples of these methods include regularised least squares    \citep[RLS;][]{christensen-dalsgaard_comparison_1990}, which has proven successful for the case of the Sun, and optimally localised averages    \citep[OLA;][]{pijpers_faster_1992,pijpers_sola_1994}, which infer the core and surface rotation rates of stars other than the Sun.

The angular momentum transport mechanism sets the rotation profile. The core-to-surface rotation ratio and the position and strength of the gradient of rotation rate can characterise the rotation profile. \citet{fellay_asteroseismology_2021} suggest that tighter constraints can be made on angular momentum transport mechanisms through more precise measures of the core-to-surface rotation ratios of post-MS stars \citep{deheuvels_seismic_2014}, and of the position and strength of a rotation rate gradient \citep{di_mauro_rotational_2018}. For example, a rotation profile with a constant rotation rate internal to the base of the convective zone (BCZ), and a decreased rotation rate that is inversely dependent on radius in the convective zone, could be indicative of angular momentum transport from deep fossil magnetic fields \citep{gough_effect_1990,kissin_rotation_2015,takahashi_modeling_2021}. This results from differential rotation being damped along poloidal field lines \citep{garaud_rotationally_2002, strugarek_magnetic_2011}. On the other hand, a steep rotational gradient near the H-burning shell of a subgiant would indicate turbulent angular momentum transport. This could be in the form of internal gravity waves \citep{pincon_can_2017}, leading to localised shallow gradients in the profile \citep{charbonnel_influence_2005} or through magneto-rotational instabilities which arise from steep angular momentum gradients \citep{spada_angular_2016,balbus_stability_1994,arlt_differential_2003,menou_magnetorotational_2006}. The gradient of the rotation profile of subgiants is not well constrained through current asteroseismic data \citep{deheuvels_seismic_2014}. 

In this work, we consider the constraints to the position of a steep rotational gradient, where we show that $\ell = 1,2$ rotational splittings may be sufficient to make valuable inferences about the rotation profiles of low-luminosity subgiants if a \TW{precise auxiliary measure of} surface rotation rate is available.
\Revision{We specifically investigate the impact of employing surface rotation periods from photometric variability owing to stellar spots \citep[e.g. those measured in][]{mcquillan_rotation_2014,garcia_rotation_2014,santos_surface_2021}.
In adopting these values, we utilise a data set that overlaps with the subset used to measure asteroseismic rotational splittings.
We deem it appropriate to employ both constraints simultaneously due to the distinct methods of measuring these quantities.}

In Section~\ref{sec:methods_res} we describe a forward model to infer rotation profile parameters given observed rotational splittings, assuming a step rotation profile. 
%Assuming a step-like rotation profile when performing inference on low-luminosity subgiants, we identify a degeneracy in the surface rotation rate and the position of the step. This implies that independent measurements of the surface rotation can place constraints on the rotation profile between the core and surface. 
We perform tests using mock data generated by three hypothetical profiles to show the differences in constraining the rotation profile with realistic \TW{independent measures of surface rotation rate }from stellar spot brightness modulations. Finally, we perform inference on the observed rotational splittings of \thestar\ with different priors  and compare the constraints on the rotation profile. The implications are discussed in Section~\ref{sec:discussion}, and summarised in Section~\ref{sec:conclusion}.

\section{Method and results}
\label{sec:methods_res}
\subsection{Rotational splittings}

Stellar oscillations can be decomposed into oscillating spherical harmonic modes. Individual modes frequencies ($\nu_{n,\ell,m}$) are characterised by their radial order ($n$), angular degree ($\ell$) and azimuthal order ($m$). 
\Revision{
Low-luminosity subgiants have much longer rotation periods (of order 10$^1$ days \citep{deheuvels_seismic_2014}) in the fast rotating core than the average oscillation period (on the order of hours \citep{aerts_asteroseismology_2010}).
We can therefore treat rotation as perturbative to the structure.
The effect of stellar rotation on oscillation mode frequencies can be approximated as perturbations to the non-rotating mode frequencies from $m=0$ to $m=-\ell$ and $m=\ell$. 
This is a widely employed approximation in the field of asteroseismic inversions of rotation \citep[e.g.][]{deheuvels_seismic_2014,deheuvels_seismic_2015,fellay_asteroseismology_2021}.
For more detail on this approach see \citet{unno_nonradial_1989,aerts_asteroseismology_2010}}.
To first order, the rotationally split oscillation frequencies are
\begin{eqnarray}
    \nu_{n,\ell,m} &=& \nu_{n,\ell,0} + \delta \nu_{n,\ell,m} \\
                   &=& \nu_{n,\ell,0} + m \ \delta \nu_{n,\ell}
\end{eqnarray}
where $\nu_{n,\ell,m}$ is the frequency of the $n,\ell,m$ mode, $\nu_{n,\ell,0}$ is the frequency in the non-rotating case, and $\delta \nu_{n,\ell,m}$ is the change in oscillation frequency due to rotation, known as the rotational splitting.
\Revision{The difference between the $m =0$ and $m \neq 0$ mode frequency is the $m$'th multiple of the $\delta \nu_{n,\ell}$ rotational splitting.
}
In practice we are usually only able to observe $\ell =1$ and $\ell=2$ modes with $m=1$ and $m=-1$ rotational splittings in low-luminosity subgiants from photometric time series data \Revision{\citep{benomar_properties_2013,deheuvels_seismic_2014}}.

Rotational splittings are scaled averages of the rotation profile. The scaling is different for each oscillation mode and is quantified using a so-called rotational kernel. The rotational kernels are inherent to the thermodynamic structure of a star \citep[see][for a derivation of these kernels]{aerts_asteroseismology_2010}. Assuming spherical symmetry, the $n,\ell$ rotational splitting is given by
   \begin{equation}
    \delta \nu_{n,\ell} (\Omega) = \beta_{n,\ell} \int^R_0 K_{n,\ell}(r) \Omega(r) \,\text{d}r
    \label{eqn:splitting}
\end{equation}
where $K_{n,\ell}$ is the rotational kernel of the $n,\ell$ mode (determined from a stellar model), $\Omega(r)$ is the scaled average, with respect to the polar axis, 1D rotation profile along the radial axis, $\beta_{n,\ell}$ is a normalisation constant, and $R$ is the outermost radius of the star. The rotational kernel, and thus scaled averaged frequency shift, changes with each oscillation mode. Changes to the rotation profile therefore result in distinct variances for each rotational splitting. Some rotation profiles are more likely to result in measurable rotational splittings than others. As a result we are able to use forward modelling to determine the set of likely rotation profiles given some observed rotational splittings.

\subsection{Forward model}
\label{sec:gen_mod}

A forward model requires a set of rotation kernels to predict rotational splittings given some profile. In this work, we use rotational kernels of the low-luminosity subgiant \thestar{}, a well-studied asteroseismic target \citep[e.g.,][]{deheuvels_seismic_2014}. Models of \thestar{} indicate that it is early in its evolution off the MS, which is supported by a relatively low core-to-surface rotation ratio. \Add{It is the earliest star known in its post-MS evolution with evidence of differential rotation.} 

We were provided with a model of \thestar{} from \citet{ball_surface-effect_2017} generated using the \software{astero} module of the Modules for Experiments in Stellar Astrophysics (\MESA{}) evolutionary code
\citep[r7624;][]{paxton_modules_2010,paxton_modules_2013,paxton_modules_2015,paxton_modules_2019}.
\Revision{The model was found by simultaneously matching the non-seismic properties of \thestar{} (T$_{\mathrm{eff}}$, $\log{g}$, and [Fe/H] in Table \ref{tab:kictab}), the global seismic quantities - 
the frequency where peak power of the Gaussian asteroseismic power envelope occurs, $\nu_\text{max}$, and the frequency spacing between consecutive radial order modes with the same angular degree, $\Delta \nu$ - and the observed mode frequencies \citep[see Tables 1 and 3 in][]{deheuvels_seismic_2014} to those predicted by the model.}
Mode frequencies were calculated using \ADIPLS{} \citep{christensen-dalsgaard_adipls_2008}, with combined surface effect corrections to the frequencies \citep{ball_new_2014, ball_surface-effect_2017}.

\Revision{
Most of the observed and best-fit model quantities agree to within 1-$\sigma$.
The exceptions are T$_\text{eff}$ and $L$  where the agreement is within 2-$\sigma$.
Our model T$_\text{eff}$ is closer to the reported T$_\text{eff}$ from the infrared flux method \citep{casagrande_absolutely_2010} (5302 $\pm$ 124 K from \citet{deheuvels_seismic_2014}), but fully consistent with the spectroscopic effective temperature.
$L$ is not included in the $\chi^2$ fitting of the model ($M$, $R$ and $L$ are outputs of the model) and the value from \citet{deheuvels_seismic_2014} was estimated from scaling relations.
This level of deviation varies with surface modelling assumptions \citep{ball_surface-effect_2017}, 
 is consistent with previous works \citep[e.g.][]{deheuvels_seismic_2014,li_asteroseismology_2020-1}, and while rotational inversions of subgiants can be dependent on model uncertainties \citep{schunker_asteroseismic_2016} this discrepancy is unlikely to affect the results of this work.}

\begin{figure}
    \includegraphics[width=0.5\textwidth]{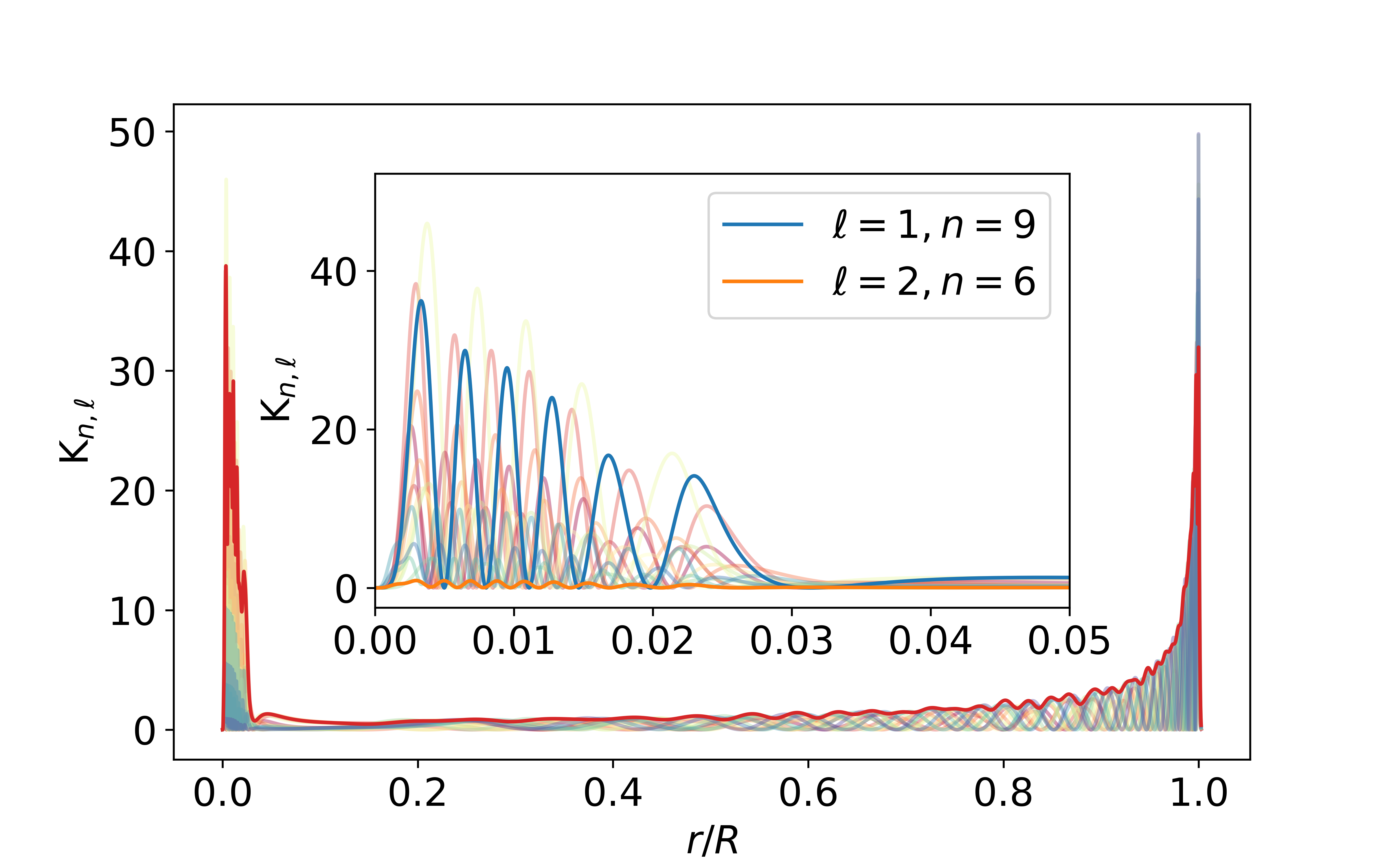}
    \caption{Rotational kernels for the best-fitting model of \thestar{}. The red curve shows three times the local standard deviation of the set of observed kernels (kernel deviation). Regions with large kernel deviation are expected to be sensitive to the rotation profile following forward modelling. In this model these regions can be seen in the H-burning shell core ($r/R<0.05$) and nearing the surface ($r/R$ > $0.8$). The inset shows the same kernels with two modes highlighted: the $\ell = 1,\ n=9$ mode (blue) is sensitive to core rotation and the $\ell = 2,\ n = 6$ (orange) is sensitive to surface rotation.}
    \label{fig:kern}
\end{figure}

We  used this model of \thestar{} to calculate the rotational kernels for all observable rotational splittings. Radial positions with both high kernel magnitude and inter-kernel variance (represented by the standard deviation of kernel magnitudes, shown in red in Figure~\ref{fig:kern}) are most sensitive to the rotation profile. For \thestar{} these regions are near the core and at $r/R$ > $0.8$.

\begin{table}
\centering
%\resizebox{\textwidth}{!}{
\begin{tabular}{|l|c|c|}
\hline
                                                  & \thestar{} & Model                  \\ \hline
$M\,(M_{\odot}$)                                   & $1.20~\pm~0.16$             & $1.293$              \\
$R\,(R_{\odot})$                                   & $2.20~\pm~0.10$             & $2.277$             \\
$L\,(L_{\odot}$)                                   & $3.25~\pm~0.45$             & $4.065$ \\
$\log{g}(\ \mathrm{cm} \mathrm{s}^{-2})$                                      & $3.83~\pm~0.04$             & $3.834$ \\
$T_{\mathrm{eff,spec}}\,(\mathrm{K})$ & $5248~\pm~130$              & $5434$     \\
$[\mathrm{Fe/H}]$ (dex)                                & $0.25~\pm~ 0.23$            & $0.06$ \\
\hline
\end{tabular}
\caption{Measured properties of \thestar{} \citep{deheuvels_seismic_2014}, and those of the best-fitting stellar model from which the rotation kernels are generated.
}
\label{tab:kictab}
\end{table}

%\Add{\citet{deheuvels_seismic_2014} established that, with current data, it is likely, not feasible to distinguish between smooth and discontinuous rotation profiles of low-luminosity subgiants. 
%We have found similar results when investigating profiles with varying gradients. 
\citet{deheuvels_seismic_2014} concluded that it is difficult to distinguish between smooth and discontinuous rotation profiles of low-luminosity subgiants using current data.
Consequently, we assume a simple form of a rotation profile $\Omega(r)$ using three parameters of the following form:
\begin{equation}
\Omega(r) = \left\{
        \begin{array}{ll}
            \Omega_c & \quad r/R \leq p \\
            \Omega_s & \quad r/R > p
        \end{array}
    \right.
    \label{eqn:step}
\end{equation}
\noindent{}where $\Omega_c$ and $\Omega_s$ describe the core and surface rotation rates respectively, and $p$ is the position of the step (in units of $r/R$). We discuss the implications of the assumed step-function form of the rotation profile further in Section \ref{sec:discussion}. Initially we will assume weak uninformed, uniform priors on the core and surface rotation rate and a uniform prior on $p$:
\begin{eqnarray}
    p &\sim& \mathcal{U}\left(0, 1\right) \\
    \Omega_s/2\pi &\sim& \mathcal{U}\left(0, 600\right) \ \mathrm{nHz} \\
    \Omega_c/2\pi &\sim& \mathcal{U}\left(0, 1000\right)\ \mathrm{nHz}\quad
\end{eqnarray}
\Revision{where $\mathcal{U}\left(x,y\right)$ denotes a uniform prior between x and y.}
We calculate the expected rotational splitting frequencies $\delta\nu_{n,l}$ for all observable $\{n,l\}$ modes given a model $\Omega(r)$ and the rotational kernels ${K}_{n,\ell}(r)$ using Eq. \ref{eqn:splitting}, and assume the observed splitting frequencies are normally distributed with a log-likelihood
\begin{equation}
    \ln \mathcal{L}( \pmb{ \delta\nu } | \Omega_c,\Omega_r,p, \mathbf{{K}(r)}, \pmb{ \sigma_{\delta\nu} } ) \propto -\frac{1}{2}\sum_{n,l}\left(\frac{\delta \nu_{n,l} - \delta \nu_{\mathrm{obs} \ n,l}}{\sigma_{\delta \nu_{n,l}}}\right)^2 
    \label{eqn:lnl}
\end{equation}
where $\delta\nu_{\mathrm{obs} \ n,l}$ is the observed rotational splitting frequency for mode $n$ and $l$, and its associated uncertainty is $\sigma_{\delta\nu_{n,l}}$.
\Revision{The vector-valued symbols $\pmb{ \delta\nu }, \mathbf{{K}(r)}, $ and $ \pmb{ \sigma_{\delta\nu} }$ indicate that the log-likelihood depends on the corresponding values for all of the observed rotational splittings.}
We constructed this model using \pymc{} \citep{salvatier_probabilistic_2016} and used the `No U-Turn Sampler' \citep{hoffman_no-u-turn_2011} to draw samples from the posterior.

\section{Results}
\label{sec:results}

\subsection{Mock data experiments with three hypothetical rotation profiles}
\label{sec:prof}

\begin{figure}
\centering
    \includegraphics[width=0.5\textwidth]{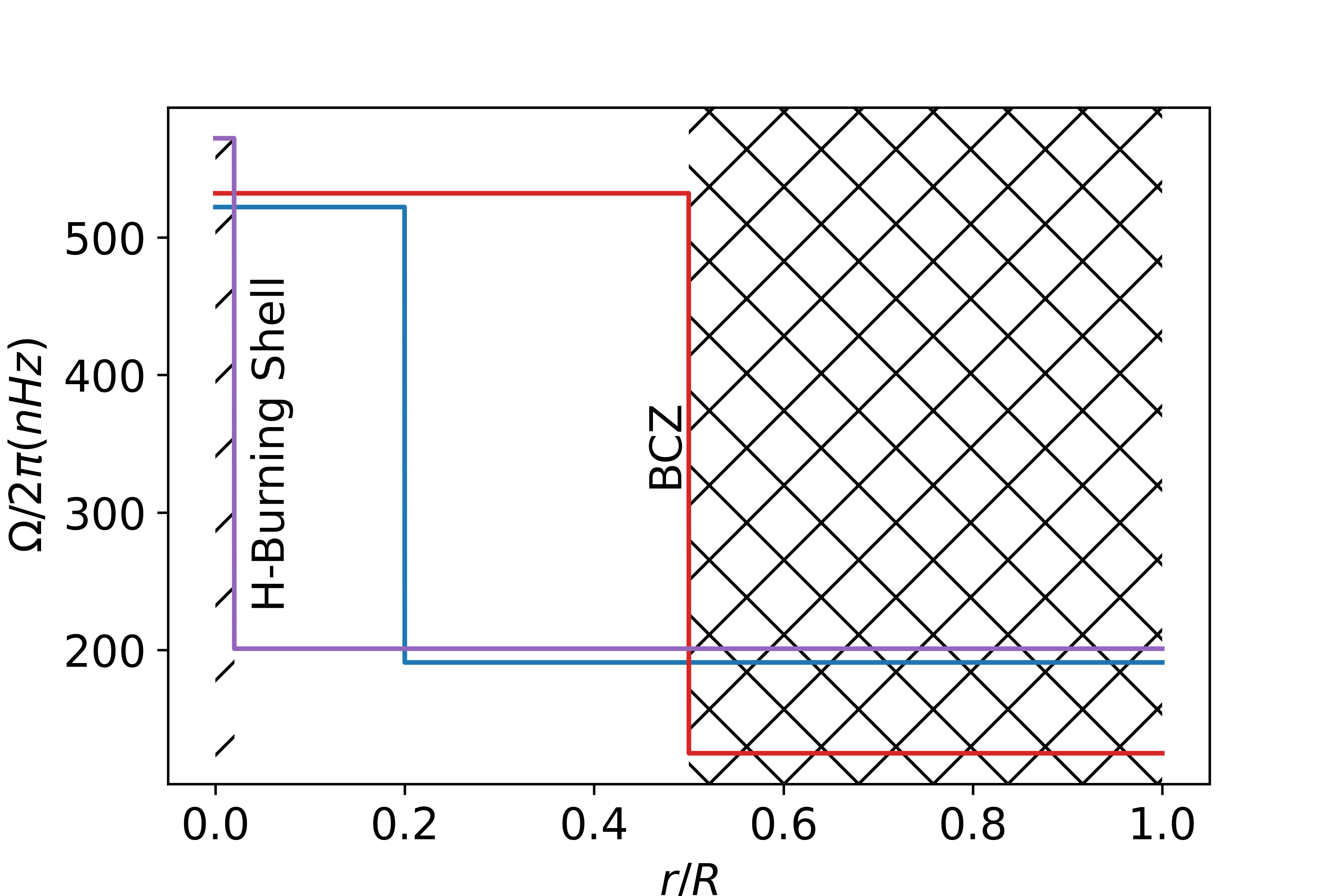}
    \caption{\Add{Three} rotation profiles used in the mock data experiments. These profiles represent the extreme consequences of different angular momentum transport mechanisms in low-luminosity post-MS stars. Cross-hatching represents the convective surface region and diagonal-hatching represents the H-burning shell. See Section~\ref{sec:prof} for descriptions of each profile.}
    \label{fig:5rotprof}
\end{figure}

We begin by generating mock data with our forward model to test the impact of \TW{independent measures of surface rotation rate}. We chose three rotation profiles with extreme differences in the position of a strong rotational gradient, which represent hypothetical angular momentum transport mechanisms that could result in step-like rotation profiles. The three step positions are: in the H-burning shell (purple); in the radiative zone (blue); and at the base of the convective zone (red) as shown in Figure \ref{fig:5rotprof}. The mock rotation profiles are motivated by, but not representative of, various angular momentum transport processes. The BCZ step rotation profile (red) is a signature of angular momentum transport by fossil magnetic fields, which results in solid body rotation in the radiative region and inverse rotation rate on radius in the convective region \citep{kissin_rotation_2015, takahashi_modeling_2021}. The H-burning step rotation profile (purple) is indicative of turbulent angular momentum transport through internal gravity waves \citep{pincon_can_2017} or magneto-rotational instabilities \citep{spada_angular_2016,balbus_stability_1994,arlt_differential_2003,menou_magnetorotational_2006} which result in a strong gradient in rotation rate close to the core. The radiative zone step profile (blue) corresponds to delocalized angular momentum transport from the core into the radiative zone and is not indicative of a specific angular momentum transport process.

% The BCZ step rotation profile (red) is a signature of angular momentum transport by fossil magnetic fields, \citep{kissin_rotation_2015, takahashi_modeling_2021}. The H-burning step rotation profile (purple) is indicative of turbulent angular momentum transport through internal gravity waves \citep{pincon_can_2017} or magneto-rotational instabilities \citep{spada_angular_2016,balbus_stability_1994,arlt_differential_2003,menou_magnetorotational_2006}. The radiative zone step profile (blue) corresponds to delocalized angular momentum transport from the core into the radiative zone and is not indicative of a specific angular momentum transport process.

The core and surface rotation rates will be realistically different for each profile/angular momentum transport process. To account for this for each profile, we fixed the step position and evaluated the log-likelihood (given the observed splittings of \thestar) at each combination of $(\Omega_c,\Omega_s)$ and set the rotation rates to those with the maximum log-likelihood. The resulting rotation profiles are shown in Figure~\ref{fig:5rotprof}: radiative-zone step (blue: $p = 0.2$), a BCZ step (red: $p = 0.5$), and the H-burning shell step (purple: $p = 0.05$).
The rotational splitting frequencies of these profiles were then calculated using  (Eq.~\ref{eqn:splitting}). 
We adopted uncertainties on those expected values given the precision of mode frequencies measured in \thestar{} (Table 3 of \citet{deheuvels_seismic_2014}). This provides us with three mock data sets to consider the rotation profile's effect on the observaations.

% \begin{figure}
% \centering
%     \includegraphics[width=0.5\textwidth]{pos_split.png}
%     \caption{\Add{Calculated rotational splittings of the H-burning shell step (purple), radiative-zone step (blue) and BCZ step (red) rotation profiles in Figure~\ref{fig:5rotprof}. The change in predicted rotational splitting frequencies among these three profiles is small compared to the 1-$\sigma$ observational uncertainties of \thestar{} (black lines).}} %as explored in Section \ref{sec:pos}}
%     \label{fig:split2}

% \end{figure}

\begin{figure*}
\centering
    \includegraphics[width=0.95\textwidth]{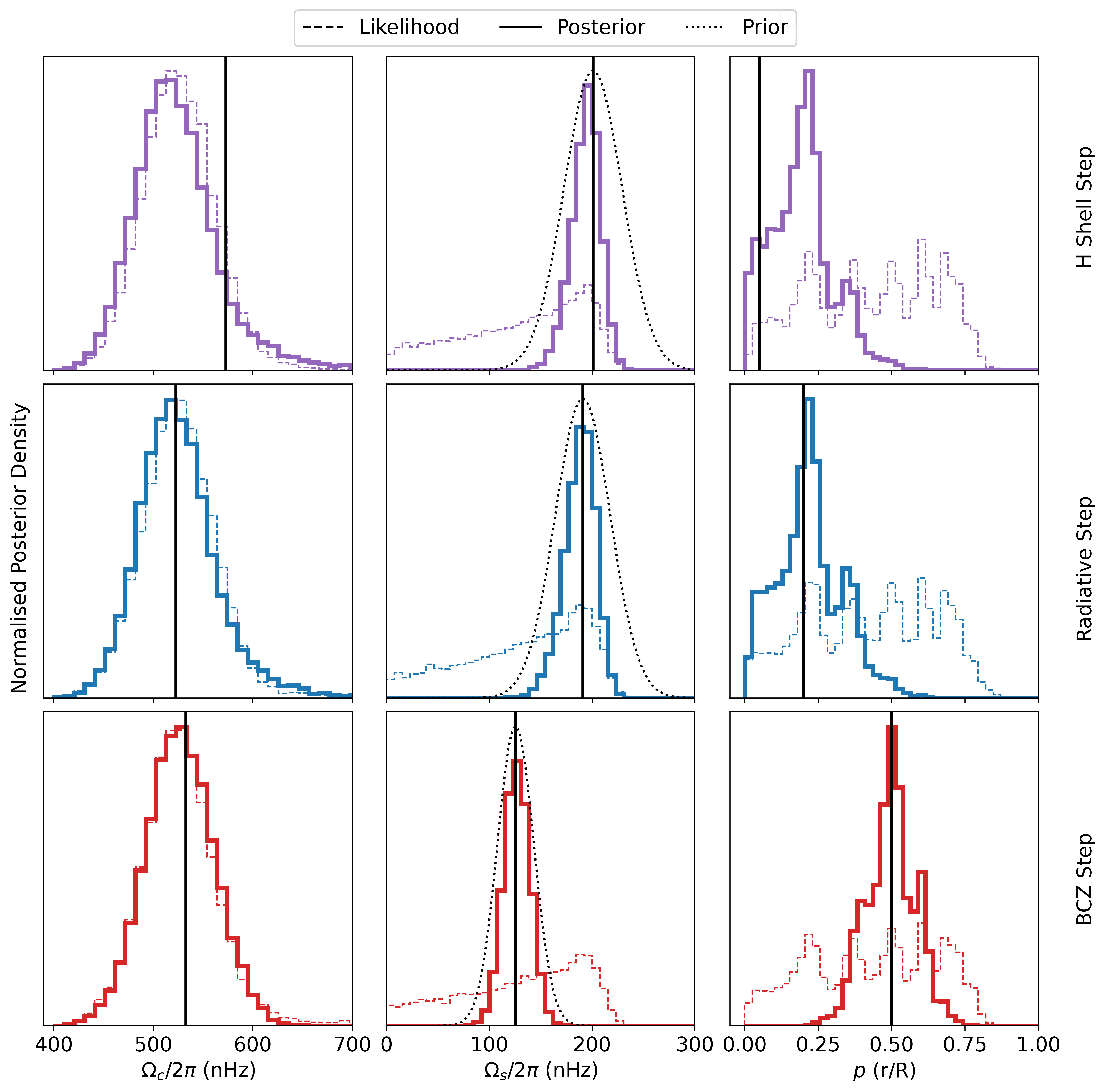}
    \caption{Normalised posterior density following sampling for each \textit{mock} rotational splitting experiment. From left to right the normalised posterior density of each of the parameters of the rotation profile model are shown: surface rotation rate ($\Omega_s/2\pi$), core rotation rate ($\Omega_c/2\pi$), and position of rotational gradient ($p$). Thin coloured dashed lines are samples with no prior on the surface rotation rate, thin dotted black lines correspond to the the introduced prior on surface rotation rate and thick coloured lines correspond to samples when the informed surface rotation prior was introduced. Vertical black lines correspond to the input values for each of the rotation profile parameters used to generate the mock rotational splittings. From top to bottom each row displays the result of sampling a different set of rotational splittings with the same colours as the rotation profiles used to generate the mock data in Figure \ref{fig:5rotprof}: the H-burning shell step (purple), radiative zone step (blue) and base of convective zone step (red).} 
    \label{fig:full_results}
\end{figure*}

We treated the mock data generated by each step profile as if it were real data and performed inference using the model and sampler described in Section~\ref{sec:gen_mod}. \TW{We performed inference twice: first with a flat prior on surface rotation and then with a Gaussian prior on surface rotation with mean equal to the injected surface rotation rate, and a standard deviation 10\% of the mean value.} Here we have chosen 10\% as representative of the average uncertainty on state-of-the-art measurements of main-sequence and subgiant stellar rotation from photometric variation \citep{santos_surface_2021}. We discuss this choice further in Section \ref{sec:discussion}. We drew 20,000 posterior samples in each case. The 1D marginalised posterior samples with a flat and informed prior are compared in Figure \ref{fig:full_results}. The full posteriors are shown in the Appendix (Figures \ref{fig:mock_posterior_005_uniform} to \ref{fig:mock_posterior_050_reject}). 

When sampling with a uniform prior on surface rotation (Figures \ref{fig:mock_posterior_005_uniform} to  \ref{fig:mock_posterior_050_uniform}), we identify multi-modality and a strong correlation between $\Omega_s$ and $p$: smaller $\Omega_s$ values tend to coincide with higher $p$ values. The inferred step position $p$ is very uncertain in all three cases, showing nearly uniform probability throughout the domain.
%This is consistent and expected from the rotational splittings shown in Figure~\ref{fig:split2}, where the variance in predictions from different step profiles is less than the measurement variance.\\

\TW{The impact of an auxiliary surface rotation rate measure on the posterior distributions is evident when comparing the normalised posterior density when using an uninformed prior (thin dashed) to the informed prior (thick solid) in Figure \ref{fig:full_results}.} When the informed prior is introduced, the degeneracy between surface rotation rate and $p$ is broken and the surface rotation rate and $p$ are better recovered. In the H Shell step (purple) and radiative step (blue) experiments, the prior has collapsed support for $p>0.4$. The introduction of the prior does not, however, allow us to differentiate between rotation profiles deeper in the star. We find that for profiles where $p\lesssim0.2$, multi-modality remains and the introduction of the prior on $\Omega_s$ increased support for $p$ closer to the core (and closer to the true value). Indeed the introduction of the informed prior for the H shell and radiative step experiments has constrained the posterior on $p$, but the $1\sigma$ range on these values overlaps significantly. We could not differentiate between these profiles using this method and state-the-art data. This is not the case for the BCZ mock data experiment. Introducing the informed surface rotation prior allows us to place more significant constraints on $p$. The posterior on $p$ was flat with an uninformed prior and the BCZ step. In contrast, with a surface rotation rate prior, the posterior is now a single peak centred at the injected value with a percentage standard deviation of the median value of about $18\%$. This illustrates the constraining power on $p$ of the independent measures of $\Omega_s$ in specific circumstances.

\subsection{\thestar}

\begin{figure}
\centering
    \includegraphics[width=0.5\textwidth]{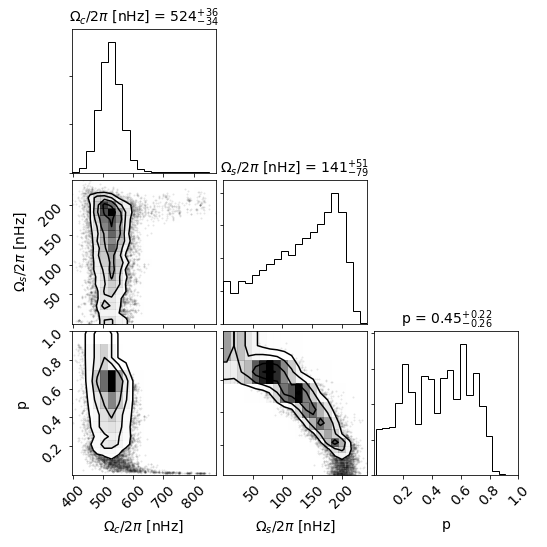}
    \caption{Posterior distributions of the core rotation rate $\Omega_c$, the surface rotation rate $\Omega_s$, and the discontinuity position $p$ (Eq.~\ref{eqn:step}) given the observed $l=\{1,2\}$ rotational splittings of \thestar{} and assuming a rotation profile with a step function.}
    \label{fig:kic-corner}
\end{figure}

\begin{figure}
\centering
    \includegraphics[width=0.5\textwidth]{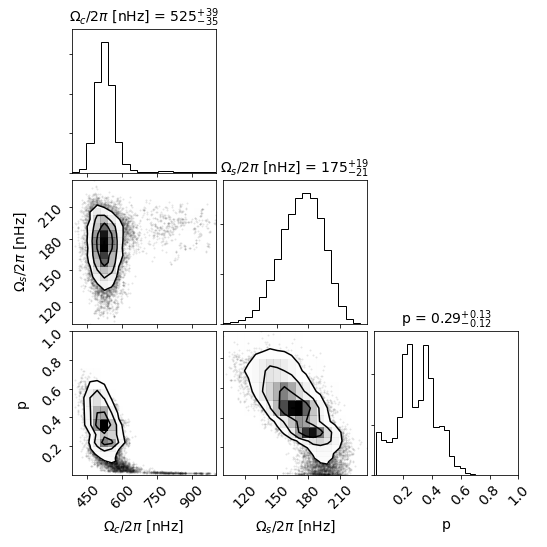}
    \caption{Posteriors on step profile parameters given the rotational splittings of \thestar\ and informed prior on surface rotation rate from \citep{garcia_rotation_2014}. }
    \label{fig:kic12408433_with_surface_rotation_prior}
\end{figure}

\TW{Our mock data experiments confirm that a realistic measure of surface rotation rate, treated as a prior during \Revision{inference} can better constrain the internal rotation profile. We now apply this method to observed rotational splittings of \thestar\ \citep{deheuvels_seismic_2014}. First with a uniform prior on surface rotation between $0$ and $600$ nHz, and then with an independent surface rotation rate \citep[measured by][from photometry]{garcia_rotation_2014} treated as a Gaussian prior on the surface rotation rate with mean 172 nHz and standard deviation 21 nHz.} The posteriors are shown in Figures \ref{fig:kic-corner} and \ref{fig:kic12408433_with_surface_rotation_prior}. The best-fitting values and credible intervals of the rotation profile parameters for each prior are shown in Table \ref{tab:results_table} and are compared to the results of optimally localised average (OLA) inversions performed in \citet{deheuvels_seismic_2014}.

\begin{table}
\resizebox{\columnwidth}{!}{%
\begin{tabular}{|c|c|c|c|}
\hline
                                     & $\Omega_c/2\pi$ (nHz)                         & $\Omega_s/2\pi$ (nHz)                        & $p$ ($r/R$)                                              \\ \hline
\citet{deheuvels_seismic_2014} (OLA)             & $532\pm79$                                    & $213\pm26$                                   & …                                                 \\ \hline
Sampling - Flat $\Omega_s$ Prior     & $524${\raisebox{0.5ex}{\tiny$^{+36}_{-34}$}}  & $141${\raisebox{0.5ex}{\tiny$^{+51}_{-79}$}}  & $0.45${\raisebox{0.5ex}{\tiny$^{+0.22}_{-0.26}$}} \\ \hline
Sampling - Informed $\Omega_s$ Prior & $525${\raisebox{0.5ex}{\tiny$^{+39}_{-35}$}} & $175${\raisebox{0.5ex}{\tiny$^{+19}_{-21}$}} & $0.29${\raisebox{0.5ex}{\tiny$^{+0.13}_{-0.12}$}} \\ \hline
\end{tabular}%
}
\caption{best-fitting rotation profile parameters given observed $\ell = 1$ and $2$ rotational splittings of \thestar\ from optimally localised averages (OLA) \citep{deheuvels_seismic_2014} and forward modelling with flat and informed ($\Omega_s/2\pi = 172 \pm 21$ nHz) priors. OLA does not place constraints on the rotation profile outside the g- and p-mode cavities in the core and at the star's surface, respectively. $p$ is more precisely measured with the introduction of the informed prior when forward modelling.}
\label{tab:results_table}
\end{table}

With a uniformed prior, the median of the posterior on the core and surface rotation rates and corresponding 16th and 84th percentile credible intervals are $\Omega_c/2\pi = 524^{+36}_{-34}\,\mathrm{nHz}$ and $\Omega_s/2\pi = {141}^{+51}_{-79}\,\mathrm{nHz}$, with the maximum a posteriori or MAP point at $\Omega_c/2\pi \approx 520$ nHz and $\Omega_s/2\pi \approx 170$ nHz. With the informed surface rotation rate prior the 16th and 84th percentile credible intervals are $\Omega_c/2\pi = 525^{+39}_{-35}\,\mathrm{nHz}$ and $\Omega_s/2\pi = {175}^{+19}_{-21}\,\mathrm{nHz}$. The inferences agree with each other, as well as the inversions performed in \citet{deheuvels_seismic_2014}. Those authors use OLA to estimate the mean rotation rate in the g-mode cavity to be $\langle\Omega_g\rangle/2\pi = 532 \pm 79\,\mathrm{nHz}$ and the mean rotation rate in the p-mode cavity to be $\langle\Omega_p\rangle/2\pi = 213 \pm 26\,\mathrm{nHz}$. Taking their $\Omega_g$ and $\Omega_p$ as analogues for $\Omega_c$ and $\Omega_s$, respectively, both sets of quantities agree within about one joint standard deviation, without considering that the \textit{mode} of the $\Omega_s$ posteriors are closer to $\langle\Omega_p\rangle$ than the median of $\Omega_s$.

Like in the mock data experiments, the \TW{independent surface rotation measure} removed support for discontinuities close to the BCZ (comparing Figures~\ref{fig:kic-corner} and \ref{fig:kic12408433_with_surface_rotation_prior}), but two peaks remain: \Add{$p \approx 0.25$ (closest to the median posterior value and a peak in the posterior when considering a flat prior on $\Omega_s$) and a much smaller peak closer to the core.}  Rotation profiles with a discontinuity located $p>0.4$ for \thestar{} is strongly disfavoured, and a discontinuity in the radiative region $~0.2 \ r/R$ is favoured. Comparing the confidence intervals when applying a flat and informed prior, we obtain a factor of approximately two increase in measurement accuracy on $p$. %$p$, and thus the rotation profile between the core and the surface of \thestar\, is more precisely measured with the introduction of the informed prior when forward modelling the rotation profile given the observed rotational splittings.

\section{Discussion}
\label{sec:discussion}

We find that a surface rotation rate measured from star spot modulations can substantially improve inferences on the rotation profiles of low-luminosity subgiants. 
This \Revision{takes advantage of} a degeneracy between $\Omega_s$ and $p$, found by performing inference using mock experiments with different discontinuity positions.

\TW{Including an independent measure of surface rotation as a prior in the analysis of \thestar\ yields increased support for step-like profiles with rotational gradients in the radiative zone and decreased support for step-like rotation profiles with $p \gtrsim 0.4$.}
%it's p \gtrsim 0.4 not 0.5 right?

\TW{In our mock data tests, we proposed that a step-like rotation profile with a rotational gradient at $0.5$ r/R (BCZ profile) is motivated by, but not representative of, angular momentum transport by fossil magnetic fields. However, the simplified step rotation profile does not accurately recreate the inverse dependency of rotation rate with radius beyond the BCZ indicative of this angular momentum transport process. Further, the signature rotation profile becomes most apparent in the red giant branch \citep{takahashi_modeling_2021, fellay_asteroseismology_2021}. Therefore, we are not proposing a lack of support for eliminating fossil magnetic fields as a possible explanation for the subgiant angular momentum transport problem.}

% Independent measures of surface rotation when sampling low-luminosity red giants with a parameterisation of the rotation profile that more accurately reflect rotation profiles arising from fossil magnetic field angular momentum transport would be required to make this claim. Red giants become g-mode dominated very early into the red giant branch \citep{gehan_core_2018, fellay_asteroseismology_2021}. As a result, the rotational splittings provide only very weak constraints to rotation close to the surface of the star. A young red-giant that expresses the effects of angular momentum transport through fossil magnetic fields would be required to investigate the claim further. As there are a limited number of post-MS with observations long enough to attempt asteroseismic inversions and the specific stellar conditions required to improve this claim, it unlikely that a candidate exists within the post-MS asteroseismic catalogue.

\TW{The auxiliary measure of surface rotation we adopt as a surface rotation rate prior }requires some discussion. The 'data' we use are rotational splitting frequencies measured from a Fourier transform of a photometric time series. The measured surface rotation rate from \citet{garcia_rotation_2014} also uses some subset of the same photometric time series data. The two sets of measurements are not strictly statistically independent, such that there is a risk that we are `stealing information'. The two approaches do use subsets of the same data set, but the methods are sufficiently distinct that we consider it reasonable to use the surface rotation rate here. 

\citet{santos_surface_2021} suggest that the average percentage uncertainty on photometric surface rotation measurements is $\sim$10\% for main-sequence and subgiant stars. The precision varies with temperature, and peaks around 6000\,K (the Kraft break). Those authors also concluded that only about 20\% of subgiants with long and precise photometric observations from \kepler{} have measurable rotation periods. Subgiants may express smaller, shorter-lived active regions, which, when combined with long rotation periods, would present a smaller observable fraction and less precise measures of the surface rotation rate when compared to main-sequence counterparts. On the other hand, observing rotational splittings in subgiants requires a long observation time, which may lend itself to a higher fraction with measurable surface rotation rates. 

It's difficult to precisely measure the surface rotation rate without using photometry. Rotational velocity ($v\sin i$) only requires a single high-resolution spectrum, but requires an estimate of the inclination angle and the star's radius, 
%but a good measurement of the inclination angle is required. 
%Spectroscopic surface velocity ($v\sin i$) does not rely on surface features; thus, there is no chance of "information stealing". As a consequence of asteroseismic analysis, a precise measure of the inclination angle is available for stars like \thestar{} and other subgiants in \citet{deheuvels_seismic_2014}. However, calculating the surface rotation rate from $v \sin i$ requires a measure of the star's radius, 
which is quickly changing in the post-MS and is strongly model dependent. 
%A limited number of subgiants and low-luminosity red giants are also observed with spectroscopic and photometric missions.

%we are unsure whether the results here necessarily generalise to other sub-giant stars and what range of stars - e.g. masses, evolutionary states, inclination, length of observation - will be most impacted by the introduction of surface rotation priors.}

We assumed a step-like (discontinuous) function for the rotation profile, consistent with other works \citep[e.g.,][]{fellay_asteroseismology_2021}. Realistically this may not reflect the true shape of the rotation profile of any low-luminosity subgiant. Asteroseismic forward modelling of the stellar rotation profile is an ill-posed problem. The observed rotational splittings alone do not uniquely determine the shape of a star's rotation profile. An assumption of the rotation profile's shape is necessary for the rotational splittings to constrain the range of possible rotation profile parameters. A much more flexible model than the step-like rotation profile could be employed. However, the inferences must be interpreted in the context of the realistic evolution of angular momentum transport. Ultimately, our work exemplifies that introducing an informed surface rotation prior further reduces the potential rotation profiles already constrained by asteroseismic data. Combining state-of-the-art asteroseismic and surface rotation data is a practical method to obtain more substantial constraints to the evolution of rotation from already available data.

In this work, we take \thestar\ as an example of a low-luminosity giant with asteroseismic observations. 
%We have found evidence that another star in the \citet{deheuvels_seismic_2014} sample (star B) expresses the degeneracy between surface rotation and $p$ (under the step-like rotation profile assumption) are yet to test other similar low-luminosity subgiants. As a result,
It is not yet clear whether the degeneracy between surface rotation and step position generalises across all subgiants of different masses, metallicities, and ages. Very few subgiants are known to have mode measurements suitable for this kind of inference. Further work is needed.

\section{Conclusions}
\label{sec:conclusion}

% Observations of subgiant core and surface rotation rates suggest excess angular momentum transport compared to state-of-the-art models of rotating stellar evolution \citep{deheuvels_seismic_2014}. Rotational splittings encode information about the rotation profile set by the underlying angular momentum transport mechanism. Previous works offer limited constraints to the shape of the rotation profile. We show that by introducing a surface rotation prior, from independent measures of the surface rotation rate, we obtain more substantial constraints to the rotation profile shape.

We investigate the impact of independent surface rotation rate priors on inference of the rotation profile of sugiant stars. We perform forward modelling of\TW{ the rotation profile given observations of rotational splittings} assuming a step function rotation profile. Under these assumptions, we identify a degeneracy between the surface rotation rate and the position of the strong gradient. Mock experiments show that the introduction of \TW{a prior on surface rotation} breaks the degeneracy, allowing us to place stronger constraints on the position of the gradient.

%We apply a prior to surface rotation from stellar spot rotational modulation to inference of the rotation profile of \thestar{}.
We introduce \TW{an independent measure of surface rotation from star spot photometric modulations as a prior when performing inference on the rotation profile of \thestar. }
We find that introducing \TW{the independent measure of surface rotation } increases the measurement precision on the position of the rotational gradient by a factor of two and that the data strongly disfavours discontinuities outside $r/R = 0.4$ in \thestar{}. Including \TW{auxiliary surface rotation measurements when performing} asteroseismic forward modelling of the rotation profile is a simple way of extending what we can learn from each star with existing data.

\section*{Acknowledgements}

We thank the anonymous referee for their thorough review of this work and for their helpful suggestions.
A.~R.~C. is supported in part by the Australian Research Council through a Discovery Early Career Researcher Award (DE190100656). Parts of this research were supported by the Australian Research Council Centre of Excellence for All Sky Astrophysics in 3 Dimensions (ASTRO 3D), through project number CE170100013. I.~M.~acknowledges support from the Australian Research Council Centre of Excellence for Gravitational Wave Discovery (OzGrav), through project number CE17010004. I.~M.~is a recipient of the Australian Research Council Future Fellowship FT190100574. W.~H.~B.~thanks the UK Science and Technology Facilities Council (STFC) for support under grant ST/R0023297/1. This work has received funding from the European Research Council (ERC) under the European Union’s Horizon 2020 research and innovation programme (CartographY GA. 804752).  Funding for the Stellar Astrophysics Centre is provided by The Danish National Research Foundation (Grant agreement no.: DNRF106). 

\section*{Data Availability}
The data and models underlying this article are available upon request to the corresponding author.

%%%%%%%%%%%%%%%%%%%%%%%%%%%%%%%%%%%%%%%%%%%%%%%%%%

%%%%%%%%%%%%%%%%%%%% REFERENCES %%%%%%%%%%%%%%%%%%

% The best way to enter references is to use BibTeX:

\bibliographystyle{mnras}
\bibliography{references} % if your bibtex file is called example.bib

\begin{thebibliography}{}
\makeatletter
\relax
\def\mn@urlcharsother{\let\do\@makeother \do\$\do\&\do\#\do\^\do\_\do\%\do\~}
\def\mn@doi{\begingroup\mn@urlcharsother \@ifnextchar [ {\mn@doi@}
  {\mn@doi@[]}}
\def\mn@doi@[#1]#2{\def\@tempa{#1}\ifx\@tempa\@empty \href
  {http://dx.doi.org/#2} {doi:#2}\else \href {http://dx.doi.org/#2} {#1}\fi
  \endgroup}
\def\mn@eprint#1#2{\mn@eprint@#1:#2::\@nil}
\def\mn@eprint@arXiv#1{\href {http://arxiv.org/abs/#1} {{\tt arXiv:#1}}}
\def\mn@eprint@dblp#1{\href {http://dblp.uni-trier.de/rec/bibtex/#1.xml}
  {dblp:#1}}
\def\mn@eprint@#1:#2:#3:#4\@nil{\def\@tempa {#1}\def\@tempb {#2}\def\@tempc
  {#3}\ifx \@tempc \@empty \let \@tempc \@tempb \let \@tempb \@tempa \fi \ifx
  \@tempb \@empty \def\@tempb {arXiv}\fi \@ifundefined
  {mn@eprint@\@tempb}{\@tempb:\@tempc}{\expandafter \expandafter \csname
  mn@eprint@\@tempb\endcsname \expandafter{\@tempc}}}

\bibitem[\protect\citeauthoryear{Aerts, Christensen-Dalsgaard  \& Kurtz}{Aerts
  et~al.}{2010}]{aerts_asteroseismology_2010}
Aerts C.,  Christensen-Dalsgaard J.,   Kurtz D.~W.,  2010, Asteroseismology,
  \mn@doi{10.1007/978-1-4020-5803-5.
}, \url {https://ui.adsabs.harvard.edu/abs/2010aste.book.....A}

\bibitem[\protect\citeauthoryear{Arlt, Hollerbach  \& Rüdiger}{Arlt
  et~al.}{2003}]{arlt_differential_2003}
Arlt R.,  Hollerbach R.,   Rüdiger G.,  2003, \mn@doi [Astronomy \&
  Astrophysics] {10.1051/0004-6361:20030251}, 401, 1087

\bibitem[\protect\citeauthoryear{Baglin}{Baglin}{2003}]{baglin_corot_2003}
Baglin A.,  2003, \mn@doi [Advances in Space Research]
  {10.1016/S0273-1177(02)00624-5}, 31, 345

\bibitem[\protect\citeauthoryear{Balbus \& Hawley}{Balbus \&
  Hawley}{1994}]{balbus_stability_1994}
Balbus S.,  Hawley J.,  1994, \mn@doi [Monthly Notices of The Royal
  Astronomical Society - MON NOTIC ROY ASTRON SOC] {10.1093/mnras/266.4.769},
  266

\bibitem[\protect\citeauthoryear{Ball \& Gizon}{Ball \&
  Gizon}{2014}]{ball_new_2014}
Ball W.~H.,  Gizon L.,  2014, \mn@doi [Astronomy \& Astrophysics]
  {10.1051/0004-6361/201424325}, 568, A123

\bibitem[\protect\citeauthoryear{Ball \& Gizon}{Ball \&
  Gizon}{2017}]{ball_surface-effect_2017}
Ball W.~H.,  Gizon L.,  2017, \mn@doi [Astronomy \& Astrophysics]
  {10.1051/0004-6361/201630260}, 600, A128

\bibitem[\protect\citeauthoryear{Benomar et~al.,}{Benomar
  et~al.}{2013}]{benomar_properties_2013}
Benomar O.,  et~al., 2013, \mn@doi [The Astrophysical Journal]
  {10.1088/0004-637X/767/2/158}, 767, 158

\bibitem[\protect\citeauthoryear{Borucki et~al.,}{Borucki
  et~al.}{2010}]{borucki_kepler_2010}
Borucki W.~J.,  et~al., 2010, \mn@doi [Science] {10.1126/science.1185402}, 327,
  977

\bibitem[\protect\citeauthoryear{Cantiello, Mankovich, Bildsten,
  Christensen-Dalsgaard  \& Paxton}{Cantiello
  et~al.}{2014}]{cantiello_angular_2014}
Cantiello M.,  Mankovich C.,  Bildsten L.,  Christensen-Dalsgaard J.,   Paxton
  B.,  2014, \mn@doi [The Astrophysical Journal] {10.1088/0004-637X/788/1/93},
  788, 93

\bibitem[\protect\citeauthoryear{Casagrande, Ramirez, Melendez, Bessell  \&
  Asplund}{Casagrande et~al.}{2010}]{casagrande_absolutely_2010}
Casagrande L.,  Ramirez I.,  Melendez J.,  Bessell M.,   Asplund M.,  2010,
  \mn@doi [Astronomy and Astrophysics] {10.1051/0004-6361/200913204}, 512, A54

\bibitem[\protect\citeauthoryear{Ceillier, Eggenberger, García  \&
  Mathis}{Ceillier et~al.}{2013}]{ceillier_understanding_2013}
Ceillier T.,  Eggenberger P.,  García R.~A.,   Mathis S.,  2013, \mn@doi
  [Astronomy and Astrophysics] {10.1051/0004-6361/201321473}, 555, A54

\bibitem[\protect\citeauthoryear{Charbonnel \& Talon}{Charbonnel \&
  Talon}{2005}]{charbonnel_influence_2005}
Charbonnel C.,  Talon S.,  2005, \mn@doi [Science] {10.1126/science.1116849},
  309, 2189

\bibitem[\protect\citeauthoryear{Christensen-Dalsgaard}{Christensen-Dalsgaard}{2008}]{christensen-dalsgaard_adipls_2008}
Christensen-Dalsgaard J.,  2008, \mn@doi [Astrophysics and Space Science]
  {10.1007/s10509-007-9689-z}, 316, 113

\bibitem[\protect\citeauthoryear{Deheuvels et~al.,}{Deheuvels
  et~al.}{2014}]{deheuvels_seismic_2014}
Deheuvels S.,  et~al., 2014, \mn@doi [Astronomy and Astrophysics]
  {10.1051/0004-6361/201322779}, 564, A27

\bibitem[\protect\citeauthoryear{Deheuvels, Ballot, Beck, Mosser, Østensen,
  García  \& Goupil}{Deheuvels et~al.}{2015}]{deheuvels_seismic_2015}
Deheuvels S.,  Ballot J.,  Beck P.~G.,  Mosser B.,  Østensen R.,  García
  R.~A.,   Goupil M.~J.,  2015, \mn@doi [Astronomy \& Astrophysics]
  {10.1051/0004-6361/201526449}, 580, A96

\bibitem[\protect\citeauthoryear{Deheuvels, Ballot, Eggenberger, Spada, Noll
  \& Hartogh}{Deheuvels et~al.}{2020}]{deheuvels_seismic_2020}
Deheuvels S.,  Ballot J.,  Eggenberger P.,  Spada F.,  Noll A.,   Hartogh J.
  W.~d.,  2020, \mn@doi [Astronomy \& Astrophysics]
  {10.1051/0004-6361/202038578}, 641, A117

\bibitem[\protect\citeauthoryear{Di~Mauro, Ventura, Corsaro  \& Moura}{Di~Mauro
  et~al.}{2018}]{di_mauro_rotational_2018}
Di~Mauro M.~P.,  Ventura R.,  Corsaro E.,   Moura B. L.~D.,  2018, \mn@doi [The
  Astrophysical Journal] {10.3847/1538-4357/aac7c4}, 862, 9

\bibitem[\protect\citeauthoryear{Eggenberger, Montalbán  \&
  Miglio}{Eggenberger et~al.}{2012}]{eggenberger_angular_2012}
Eggenberger P.,  Montalbán J.,   Miglio A.,  2012, \mn@doi [Astronomy \&
  Astrophysics] {10.1051/0004-6361/201219729}, 544, L4

\bibitem[\protect\citeauthoryear{Eggenberger et~al.,}{Eggenberger
  et~al.}{2019}]{eggenberger_asteroseismology_2019}
Eggenberger P.,  et~al., 2019, \mn@doi [Astronomy \& Astrophysics]
  {10.1051/0004-6361/201833447}, 621, A66

\bibitem[\protect\citeauthoryear{Fellay, Buldgen, Eggenberger, Khan, Salmon,
  Miglio  \& Montalbán}{Fellay et~al.}{2021}]{fellay_asteroseismology_2021}
Fellay L.,  Buldgen G.,  Eggenberger P.,  Khan S.,  Salmon S. J. A.~J.,  Miglio
  A.,   Montalbán J.,  2021, \mn@doi [Astronomy \& Astrophysics]
  {10.1051/0004-6361/202140518}, 654, A133

\bibitem[\protect\citeauthoryear{Fuller, Cantiello, Stello, Garcia  \&
  Bildsten}{Fuller et~al.}{2015}]{fuller_asteroseismology_2015}
Fuller J.,  Cantiello M.,  Stello D.,  Garcia R.~A.,   Bildsten L.,  2015,
  \mn@doi [Science] {10.1126/science.aac6933}, 350, 423

\bibitem[\protect\citeauthoryear{Garaud}{Garaud}{2002}]{garaud_rotationally_2002}
Garaud P.,  2002, \mn@doi [Monthly Notices of the Royal Astronomical Society]
  {10.1046/j.1365-8711.2002.05662.x}, 335, 707

\bibitem[\protect\citeauthoryear{Garcia et~al.,}{Garcia
  et~al.}{2014}]{garcia_rotation_2014}
Garcia R.~A.,  et~al., 2014, \mn@doi [Astronomy \& Astrophysics]
  {10.1051/0004-6361/201423888}, 572, A34

\bibitem[\protect\citeauthoryear{Gehan, Mosser, Michel, Samadi  \&
  Kallinger}{Gehan et~al.}{2018}]{gehan_core_2018}
Gehan C.,  Mosser B.,  Michel E.,  Samadi R.,   Kallinger T.,  2018, \mn@doi
  [Astronomy \& Astrophysics] {10.1051/0004-6361/201832822}, 616, A24

\bibitem[\protect\citeauthoryear{Gough}{Gough}{2015}]{gough_glimpses_2015}
Gough D.~O.,  2015, \mn@doi [Space Science Reviews]
  {10.1007/s11214-015-0159-6}, 196, 15

\bibitem[\protect\citeauthoryear{Gough \& Thompson}{Gough \&
  Thompson}{1990}]{gough_effect_1990}
Gough D.~O.,  Thompson M.~J.,  1990, \mn@doi [Monthly Notices of the Royal
  Astronomical Society] {10.1093/mnras/242.1.25}, 242, 25

\bibitem[\protect\citeauthoryear{Heger}{Heger}{1998}]{heger_presupernova_1998}
Heger A.,  1998, PhD thesis, Technical University of Munich

\bibitem[\protect\citeauthoryear{Hermes, Kawaler, Bischoff-Kim, Provencal,
  Dunlap  \& Clemens}{Hermes et~al.}{2017}]{hermes_deep_2017}
Hermes J.~J.,  Kawaler S.~D.,  Bischoff-Kim A.,  Provencal J.~L.,  Dunlap
  B.~H.,   Clemens J.~C.,  2017, \mn@doi [The Astrophysical Journal]
  {10.3847/1538-4357/835/2/277}, 835, 277

\bibitem[\protect\citeauthoryear{Hoffman \& Gelman}{Hoffman \&
  Gelman}{2011}]{hoffman_no-u-turn_2011}
Hoffman M.~D.,  Gelman A.,  2011, The {No}-{U}-{Turn} {Sampler}: {Adaptively}
  {Setting} {Path} {Lengths} in {Hamiltonian} {Monte} {Carlo},
  \mn@doi{10.48550/arXiv.1111.4246}, \url {http://arxiv.org/abs/1111.4246}

\bibitem[\protect\citeauthoryear{Kissin \& Thompson}{Kissin \&
  Thompson}{2015}]{kissin_rotation_2015}
Kissin Y.,  Thompson C.,  2015, \mn@doi [The Astrophysical Journal]
  {10.1088/0004-637X/808/1/35}, 808, 35

\bibitem[\protect\citeauthoryear{Li, Bedding, Li, Bi, Stello, Zhou  \&
  White}{Li et~al.}{2020}]{li_asteroseismology_2020-1}
Li Y.,  Bedding T.~R.,  Li T.,  Bi S.,  Stello D.,  Zhou Y.,   White T.~R.,
  2020, \mn@doi [Monthly Notices of the Royal Astronomical Society]
  {10.1093/mnras/staa1335}, 495, 2363

\bibitem[\protect\citeauthoryear{Maeder \& Meynet}{Maeder \&
  Meynet}{2000}]{maeder_evolution_2000}
Maeder A.,  Meynet G.,  2000, \mn@doi [Annual Review of Astronomy and
  Astrophysics] {10.1146/annurev.astro.38.1.143}, 38, 143

\bibitem[\protect\citeauthoryear{Marques et~al.,}{Marques
  et~al.}{2013}]{marques_seismic_2013}
Marques J.~P.,  et~al., 2013, \mn@doi [Astronomy \& Astrophysics]
  {10.1051/0004-6361/201220211}, 549, A74

\bibitem[\protect\citeauthoryear{McQuillan, Mazeh  \& Aigrain}{McQuillan
  et~al.}{2014}]{mcquillan_rotation_2014}
McQuillan A.,  Mazeh T.,   Aigrain S.,  2014, \mn@doi [The Astrophysical
  Journal Supplement Series] {10.1088/0067-0049/211/2/24}, 211, 24

\bibitem[\protect\citeauthoryear{Menou \& Mer}{Menou \&
  Mer}{2006}]{menou_magnetorotational_2006}
Menou K.,  Mer J.~L.,  2006, \mn@doi [The Astrophysical Journal]
  {10.1086/507022}, 650, 1208

\bibitem[\protect\citeauthoryear{Noll, Deheuvels  \& Ballot}{Noll
  et~al.}{2021}]{noll_probing_2021}
Noll A.,  Deheuvels S.,   Ballot J.,  2021, \mn@doi [Astronomy \& Astrophysics]
  {10.1051/0004-6361/202040055}, 647

\bibitem[\protect\citeauthoryear{Ouazzani, Marques, Goupil, Christophe, Antoci
  \& Salmon}{Ouazzani et~al.}{2018}]{ouazzani_gamma_2018}
Ouazzani R.-M.,  Marques J.~P.,  Goupil M.-J.,  Christophe S.,  Antoci V.,
  Salmon S. J. A.~J.,  2018, arXiv e-prints, 1801, arXiv:1801.09228

\bibitem[\protect\citeauthoryear{Paxton, Bildsten, Dotter, Herwig, Lesaffre  \&
  Timmes}{Paxton et~al.}{2010}]{paxton_modules_2010}
Paxton B.,  Bildsten L.,  Dotter A.,  Herwig F.,  Lesaffre P.,   Timmes F.,
  2010, The Astrophysical Journal Supplement Series, 192, 3

\bibitem[\protect\citeauthoryear{Paxton et~al.,}{Paxton
  et~al.}{2013}]{paxton_modules_2013}
Paxton B.,  et~al., 2013, The Astrophysical Journal Supplement Series, 208, 4

\bibitem[\protect\citeauthoryear{Paxton et~al.,}{Paxton
  et~al.}{2015}]{paxton_modules_2015}
Paxton B.,  et~al., 2015, The Astrophysical Journal Supplement Series, 220, 15

\bibitem[\protect\citeauthoryear{Paxton et~al.,}{Paxton
  et~al.}{2019}]{paxton_modules_2019}
Paxton B.,  et~al., 2019, The Astrophysical Journal Supplement Series, 243, 10

\bibitem[\protect\citeauthoryear{Pinçon, Belkacem, Goupil  \& Marques}{Pinçon
  et~al.}{2017}]{pincon_can_2017}
Pinçon C.,  Belkacem K.,  Goupil M.~J.,   Marques J.~P.,  2017, \mn@doi
  [Astronomy \& Astrophysics] {10.1051/0004-6361/201730998}, 605, A31

\bibitem[\protect\citeauthoryear{Salvatier, Wiecki  \& Fonnesbeck}{Salvatier
  et~al.}{2016}]{salvatier_probabilistic_2016}
Salvatier J.,  Wiecki T.~V.,   Fonnesbeck C.,  2016, \mn@doi [PeerJ Computer
  Science] {10.7717/peerj-cs.55}, 2, e55

\bibitem[\protect\citeauthoryear{Santos, Breton, Mathur  \& García}{Santos
  et~al.}{2021}]{santos_surface_2021}
Santos A. R.~G.,  Breton S.~N.,  Mathur S.,   García R.~A.,  2021, \mn@doi
  [The Astrophysical Journal Supplement Series] {10.3847/1538-4365/ac033f},
  255, 17

\bibitem[\protect\citeauthoryear{Schunker, Schou, Ball, Nielsen  \&
  Gizon}{Schunker et~al.}{2016}]{schunker_asteroseismic_2016}
Schunker H.,  Schou J.,  Ball W.~H.,  Nielsen M.~B.,   Gizon L.,  2016,
  Astronomy \& Astrophysics, 586, A79

\bibitem[\protect\citeauthoryear{Spada, Gellert, Arlt  \& Deheuvels}{Spada
  et~al.}{2016}]{spada_angular_2016}
Spada F.,  Gellert M.,  Arlt R.,   Deheuvels S.,  2016, \mn@doi [Astronomy \&
  Astrophysics] {10.1051/0004-6361/201527591}, 589, A23

\bibitem[\protect\citeauthoryear{Strugarek, Brun  \& Zahn}{Strugarek
  et~al.}{2011}]{strugarek_magnetic_2011}
Strugarek A.,  Brun S.,   Zahn J.,  2011, \mn@doi [Astronomy \& Astrophysics -
  ASTRON ASTROPHYS] {10.1051/0004-6361/201116518}, 532

\bibitem[\protect\citeauthoryear{Takahashi \& Langer}{Takahashi \&
  Langer}{2021}]{takahashi_modeling_2021}
Takahashi K.,  Langer N.,  2021, \mn@doi [Astronomy \& Astrophysics]
  {10.1051/0004-6361/202039253}, 646, A19

\bibitem[\protect\citeauthoryear{Unno, Osaki, Ando, Saio  \& Shibahashi}{Unno
  et~al.}{1989}]{unno_nonradial_1989}
Unno W.,  Osaki Y.,  Ando H.,  Saio H.,   Shibahashi H.,  1989, Nonradial
  oscillations of stars.
\url {https://ui.adsabs.harvard.edu/abs/1989nos..book.....U}

\makeatother
\end{thebibliography}

% Alternatively you could enter them by hand, like this:
% This method is tedious and prone to error if you have lots of references
%\begin{thebibliography}{99}
%\bibitem[\protect\citeauthoryear{Author}{2012}]{Author2012}
%Author A.~N., 2013, Journal of Improbable Astronomy, 1, 1
%\bibitem[\protect\citeauthoryear{Others}{2013}]{Others2013}
%Others S., 2012, Journal of Interesting Stuff, 17, 198
%\end{thebibliography}

%%%%%%%%%%%%%%%%%%%%%%%%%%%%%%%%%%%%%%%%%%%%%%%%%%

%%%%%%%%%%%%%%%%% APPENDICES %%%%%%%%%%%%%%%%%%%%%

\appendix
\section{Sampling Results}
Here we provide the posteriors following sampling of each set of rotational splittings.

\begin{figure}
\centering
    \includegraphics[width=0.5\textwidth]{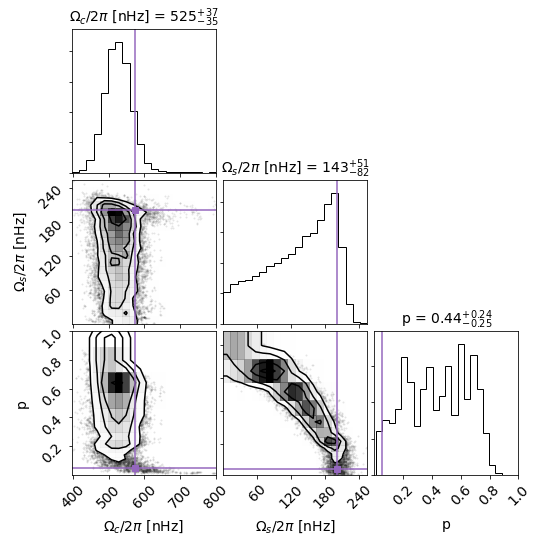}
    \caption{
    Posterior distributions using mock data generated with a step function aligned with the H burning shell ($r/R = 0.05$, purple profile in Figure~\ref{fig:5rotprof}). True values are indicated in purple. There is considerable multi-modality and degeneracy present.}
    \label{fig:mock_posterior_005_uniform}
\end{figure}

\begin{figure}
\centering
    \includegraphics[width=0.5\textwidth]{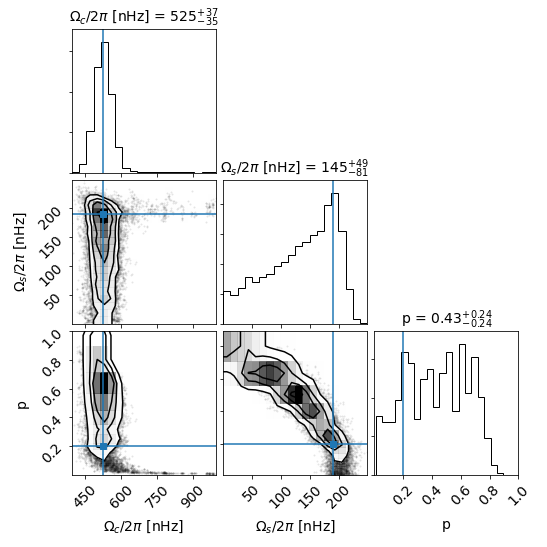}
    \caption{Posterior distributions using mock data generated with a step function in the radiative region ($r/R = 0.2$, blue profile in Figure~\ref{fig:5rotprof}) and realistic uncertainties. True values in blue.}
    \label{fig:mock_posterior_020_uniform}
\end{figure}
\begin{figure}
\centering
    \includegraphics[width=0.5\textwidth]{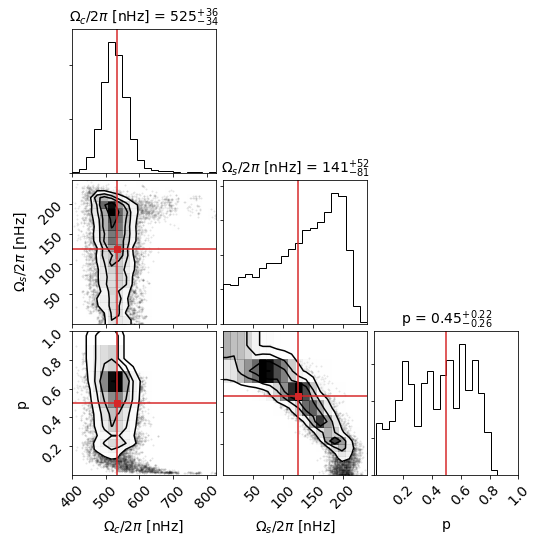}
    \caption{Posterior distributions using mock data generated with a step function at the BCZ ($r/R = 0.5$; red profile in Figure~\ref{fig:5rotprof}), and realistic uncertainties. True values in red.}
    \label{fig:mock_posterior_050_uniform}
\end{figure}

\begin{figure}
\centering
    \includegraphics[width=0.5\textwidth]{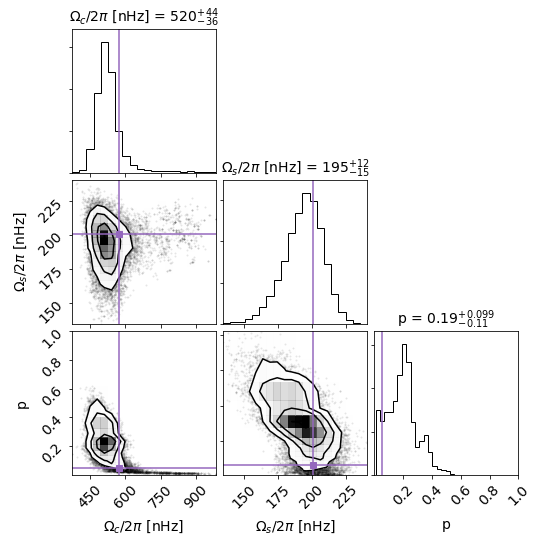}
    \caption{Posterior distributions using mock data generated with a step profile at the g-mode cavity ($r/R = 0.05$; purple profile in Figure~\ref{fig:5rotprof}), with realistic uncertainties, and a 10\% prior on surface rotation $\Omega_s$. There is still degeneracy between $p$ and the rotation parameters (e.g., Figure~\ref{fig:mock_posterior_005_uniform}), but the prior has collapsed all other modes.}
    \label{fig:mock_posterior_005_reject}
\end{figure}

\begin{figure}
\centering
    \includegraphics[width=0.5\textwidth]{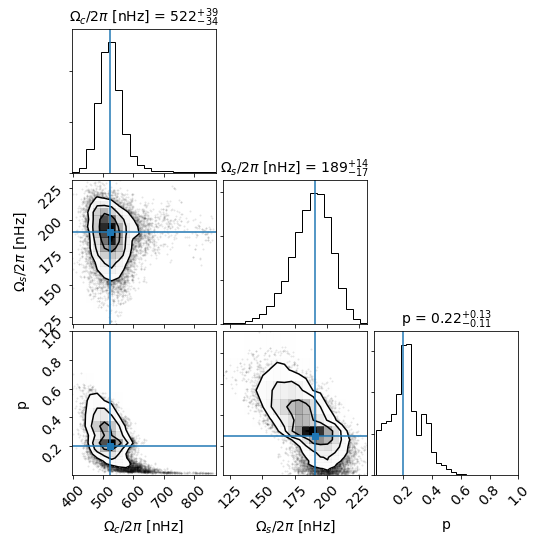}
    \caption{Posterior distributions using mock data generated with a step profile in the radiative region ($r/R = 0.20$; blue profile in Figure~\ref{fig:5rotprof}), with realistic uncertainties, and a 10\% prior on surface rotation $\Omega_s$ (compare with Figure~\ref{fig:mock_posterior_020_uniform}).}
    \label{fig:mock_posterior_020_reject}
\end{figure}

\begin{figure}
\centering
    \includegraphics[width=0.5\textwidth]{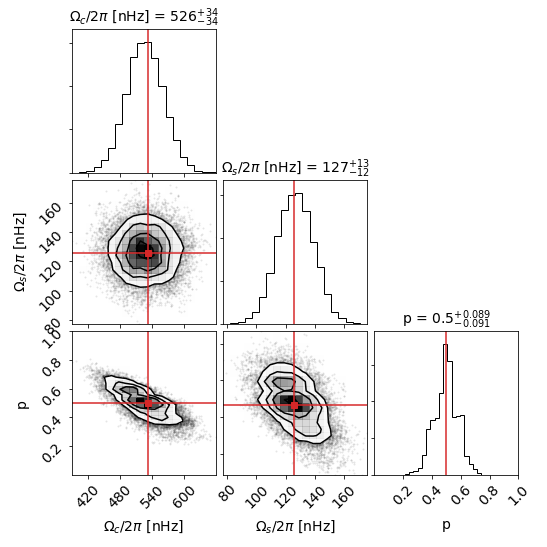}
    \caption{Posterior distributions using mock data generated with a step profile at the BCZ ($r/R = 0.50$; red profile in Figure~\ref{fig:5rotprof}), with realistic uncertainties, and a 10\% prior on surface rotation $\Omega_s$ (compare with Figure~\ref{fig:mock_posterior_050_uniform}).}
    \label{fig:mock_posterior_050_reject}
\end{figure}
% Don't change these lines
\bsp	% typesetting comment
\label{lastpage}
\end{document}